\documentclass[11pt]{article}
\usepackage{amssymb,amsmath,amsfonts}
\usepackage{graphicx}
\usepackage{graphics}
\usepackage{eepic,epsfig}

\begin{document}

\title{Fermionic Casimir densities in toroidally compactified spacetimes
with applications to nanotubes}
\author{ S. Bellucci$^{1}$\thanks{%
E-mail: bellucci@lnf.infn.it } and A. A. Saharian$^{2}$\thanks{%
E-mail: saharian@ictp.it } \\
\textit{$^1$ INFN, Laboratori Nazionali di Frascati,}\\
\textit{Via Enrico Fermi 40,00044 Frascati, Italy} \vspace{0.3cm}\\
\textit{$^2$ Department of Physics, Yerevan State University,}\\
\textit{1 Alex Manoogian Street, 0025 Yerevan, Armenia }}
\maketitle

\begin{abstract}
Fermionic condensate and the vacuum expectation values of the
energy-momentum tensor are investigated for a massive spinor fields in
higher-dimensional spacetimes with an arbitrary number of toroidally
compactified spatial dimensions. By using the Abel-Plana summation formula
and the zeta function technique we present the vacuum expectation values in
two different forms. Applications of the general formulae to cylindrical and
toroidal carbon nanotubes are given. We show that the topological Casimir
energy is positive for metallic cylindrical nanotubes and is negative for
semiconducting ones. The toroidal compactification of a cylindrical nanotube
along its axis increases the Casimir energy for metallic-type (periodic)
boundary conditions along its axis and decreases the Casimir energy for the
semiconducting-type compactifications.
\end{abstract}

\bigskip


\bigskip

\section{Introduction}

Many of high energy theories of fundamental physics, including supergravity
and superstring theories, are formulated in spacetimes having compact
spatial dimensions. From an inflationary point of view universes with
compact dimensions, under certain conditions, should be considered a rule
rather than an exception \cite{Lind04}. The models of a compact universe
with non-trivial topology may play an important role by providing proper
initial conditions for inflation. There are many reasons to expect that in
string theory the most natural topology for the universe is that of a flat
compact three-manifold \cite{McIn04}. The quantum creation of the universe
having toroidal spatial topology is discussed in \cite{Zeld84} and in
references \cite{Gonc85} within the framework of various supergravity
theories. An interesting application of the quantum field theoretical models
with non-trivial topology of spatial dimensions recently appeared in
nanophysics \cite{Sait98}. In a sheet of hexagons from the graphite
structure, known as graphene, the long-wavelength description of the
electronic states can be formulated in terms of the Dirac-like theory of
massless spinors in 3-dimensional spacetime with the Fermi velocity playing
the role of speed of light (see, e.g., Refs. \cite{Seme84}). Single-walled
carbon nanotubes are generated by rolling up a graphene sheet to form a
cylinder and the background spacetime for the corresponding Dirac-like
theory has topology $R^{2}\times S^{1}$. Compactifying the direction along
the cylinder axis we obtain another class of graphene made structures called
toroidal carbon nanotubes with the background topology $R^{1}\times
(S^{1})^{2}$.

The compactification of spatial dimensions leads to a number of interesting
quantum field theoretical effects which include instabilities in interacting
field theories \cite{Ford80a}, topological mass generation \cite{Ford79},
symmetry breaking \cite{Toms80b}. In the case of nontrivial topology the
boundary conditions imposed on fields give rise to the modification of the
spectrum for vacuum fluctuations and, as a result, to the Casimir-type
contributions in the vacuum expectation values of physical observables (for
the topological Casimir effect and its role in cosmology see \cite{Most97}-%
\cite{Duff86} and references therein). The Casimir effect is common to all
systems characterized by fluctuating quantities and has important
implications on all scales, from cosmological to subnuclear. In the
Kaluza-Klein-type models this effect has been used as a stabilization
mechanism for moduli fields which parametrize the size and the shape of the
extra dimensions. The Casimir energy can also serve as a model for dark
energy needed for the explanation of the present accelerated expansion of
the universe (see \cite{Milt03} and references therein). In addition to its
fundamental interest the Casimir effect also plays an important role in the
fabrication and operation of nano- and micro-scale mechanical systems (see,
for instance, \cite{CasNano}) and has become an increasingly popular topic
in quantum field theory.

The effects of the toroidal compactification of spatial dimensions on the
properties of quantum vacuum for various spin fields have been discussed by
several authors (see, for instance, \cite{Gonc85},\cite{Most97}-\cite{Duff86}%
, \cite{CasTor,Saha08} and references therein). In the present paper, we
investigate one-loop quantum effects arising from vacuum fluctuations of a
massive fermionic field on background of higher-dimensional spacetimes with
an arbitrary number of toroidally compactified spatial dimensions. We will
assume generalized periodicity conditions along compactified dimensions with
arbitrary phases. Important quantities that characterize the quantum
fluctuations are the fermionic condensate and the expectation value of the
energy-momentum tensor. In the next section, by using the Abel-Plana
summation formula, we derive a recurrence formula relating the fermionic
condensates in topologies $R^{p}\times (S^{1})^{q}$ and $R^{p+1}\times
(S^{1})^{q-1}$. An alternative expression for the topological part in the
fermionic condensate is obtained by using the zeta function technique. In
section \ref{sec:EMT} we consider the corresponding formulae for the vacuum
expectation values of the energy-momentum tensor. In section \ref%
{sec:Nanotubes} we give applications of general formulae to the Casimir
effect for electrons in a carbon nanotube within the framework of
3-dimensional Dirac-like model. The main results of the paper are summarized
in section \ref{sec:Conclusion}. In appendix we show the equivalence of two
representations for the vacuum expectation values obtained by the Abel-Plana
summation formula and by the zeta function method.

\section{Fermionic condensate}

\label{sec:WF}

We consider a quantum fermionic field on background of $(D+1)$-dimensional
flat spacetime with spatial topology $\mathrm{R}^{p}\times (\mathrm{S}%
^{1})^{q}$, $p+q=D$. The corresponding line element has the form
\begin{equation}
ds^{2}=dt^{2}-\sum_{l=1}^{D}(dz^{l})^{2},  \label{ds2}
\end{equation}%
where $-\infty <z^{l}<\infty $, $l=1,\ldots ,p$, and $0\leqslant
z^{l}\leqslant L_{l}$ for $l=p+1,\ldots ,D$. The dynamics of the field is
governed by the Dirac equation
\begin{equation}
i\gamma ^{\mu }\partial _{\mu }\psi -m\psi =0\ .  \label{Direq}
\end{equation}%
In the $(D+1)$-dimensional spacetime the Dirac matrices are $N\times N$
matrices with $N=2^{[(D+1)/2]}$, where the square brackets mean the integer
part of the enclosed expression. We will assume that these matrices are
given in the chiral representation:
\begin{equation}
\gamma ^{0}=\left(
\begin{array}{cc}
1 & 0 \\
0 & -1%
\end{array}%
\right) ,\;\gamma ^{\mu }=\left(
\begin{array}{cc}
0 & \sigma _{\mu } \\
-\sigma _{\mu }^{+} & 0%
\end{array}%
\right) ,\;\mu =1,2,\ldots ,D,  \label{gam0mu}
\end{equation}%
with the relation $\sigma _{\mu }\sigma _{\nu }^{+}+\sigma _{\nu }\sigma
_{\mu }^{+}=2\delta _{\mu \nu }$. For example, in $D=4$ the first four
matrices $\gamma ^{\mu }$, $\mu =0,1,2,3$, can be taken the same as the
corresponding matrices in 4-dimensional spacetime and $\gamma ^{4}=\gamma
^{0}\gamma ^{1}\gamma ^{2}\gamma ^{3}$. In this case $\sigma _{1},\sigma
_{2},\sigma _{3}$ are the standard Pauli matrices and%
\begin{equation}
\sigma _{4}=\left(
\begin{array}{cc}
0 & -i \\
-i & 0%
\end{array}%
\right) .  \label{sigma4}
\end{equation}%
Note that, unlike to the Pauli matrices, $\sigma _{4}$ is antihermitian.

In this paper we are interested in the effects of non-trivial topology on
the vacuum expectation values (VEVs) of the energy-momentum tensor and the
fermionic condensate assuming that along the compactified dimensions the
field obeys the boundary conditions (no summation over $l=p+1,\ldots ,D$)%
\begin{equation}
\psi (t,\mathbf{z}_{p},\mathbf{z}_{q}+L_{l}\mathbf{e}_{l})=e^{2\pi i\alpha
_{l}}\psi (t,\mathbf{z}_{p},\mathbf{z}_{q}),  \label{BC}
\end{equation}%
with constant phases $\alpha _{l}$. In (\ref{BC}), $\mathbf{z}%
_{p}=(z^{1},\ldots ,z^{p})$ and $\mathbf{z}_{q}=(z^{p+1},\ldots ,z^{D})$
denote the coordinates along uncompactified and compactified dimensions
respectively, $\mathbf{e}_{l}$\ is the unit vector along the direction of
the coordinate $z^{l}$. First we consider the fermionic condensate.

For the topology under consideration the fermionic condensate $\langle 0|%
\bar{\psi}\psi |0\rangle $ (with $|0\rangle $ being the amplitude for the
vacuum state) we will denote by $\langle \bar{\psi}\psi \rangle _{p,q}$. We
expand the field operator in terms of the complete set of positive and
negative frequency eigenfunctions $\{\psi _{\beta }^{(+)},\psi _{\beta
}^{(-)}\}$:
\begin{equation}
\hat{\psi}=\sum_{\beta }[\hat{a}_{\beta }\psi _{\beta }^{(+)}+\hat{b}_{\beta
}^{+}\psi _{\beta }^{(-)}],  \label{operatorexp}
\end{equation}%
where $\hat{a}_{\beta }$ is the annihilation operator for particles, and $%
\hat{b}_{\beta }^{+}$ is the creation operator for antiparticles. By using
the commutation relations for these operators, the condensate is presented
in the form of the mode-sum%
\begin{equation}
\langle \bar{\psi}\psi \rangle _{p,q}=\sum_{\beta }\bar{\psi}_{\beta
}^{(-)}(x)\psi _{\beta }^{(-)}(x).  \label{FCsum}
\end{equation}%
In order to evaluate the condensate by this formula we need the explicit
form of the eigenfunctions satisfying the boundary conditions (\ref{BC}).

In accordance with the problem symmetry the dependence of these functions on
the spacetime coordinates can be taken in the plane-wave form $e^{i\mathbf{k}%
\cdot \mathbf{r}-i\omega t}$, $\omega =\sqrt{k^{2}+m^{2}}$, with the wave
vector $\mathbf{k}$. From the Dirac equation we find
\begin{eqnarray}
\psi _{\beta }^{(+)} &=&\frac{e^{i\mathbf{k}\cdot \mathbf{r}-i\omega t}}{%
(2^{p+1}\pi ^{p}V_{q}\omega )^{1/2}}\left(
\begin{array}{c}
w_{\sigma }^{(+)}\sqrt{\omega +m} \\
(\mathbf{n}\cdot \boldsymbol{\sigma})w_{\sigma }^{(+)}\sqrt{\omega -m}%
\end{array}%
\right) ,  \label{Funcpos} \\
\psi _{\beta }^{(-)} &=&\frac{e^{-i\mathbf{k}\cdot \mathbf{r}+i\omega t}}{%
(2^{p+1}\pi ^{p}V_{q}\omega )^{1/2}}\left(
\begin{array}{c}
(\mathbf{n}\cdot \boldsymbol{\sigma})w_{\sigma }^{(-)}\sqrt{\omega -m} \\
w_{\sigma }^{(-)}\sqrt{\omega +m}%
\end{array}%
\right) ,  \label{Funcneg}
\end{eqnarray}%
where $\beta =(\mathbf{k},\sigma )$, $\mathbf{n}=\mathbf{k}/k$, and $%
\boldsymbol{\sigma}=(\sigma _{1},\sigma _{2},\ldots ,\sigma _{D})$, $%
V_{q}=L_{p+1}\cdots L_{D}$ is the volume of the compactified subspace. In
these expressions $w_{\sigma }^{(+)}$, $\sigma =1,\ldots ,N/2$, are
one-column matrices having $N/2$ rows with the elements $w_{l}^{(\sigma
)}=\delta _{l\sigma }$, and $w_{\sigma }^{(-)}=iw_{\sigma }^{(+)}$. The
eigenfunctions (\ref{Funcpos}), (\ref{Funcneg}) are normalized in accordance
with the condition%
\begin{equation}
\int d^{D}x\,\psi _{\beta }^{(\pm )+}\psi _{\beta ^{\prime }}^{(\pm
)}=\delta _{\beta \beta ^{\prime }}.  \label{normaliz}
\end{equation}%
In the discussion below we will decompose the wave vector into components
along the uncompactified and compactified dimensions: $\mathbf{k}=(\mathbf{k}%
_{p},\mathbf{k}_{q})$, $k=\sqrt{\mathbf{k}_{p}^{2}+\mathbf{k}_{q}^{2}}$. The
eigenvalues for the components along the compactified dimensions are
determined from the boundary conditions (\ref{BC}):%
\begin{equation}
\mathbf{k}_{q}=(2\pi (n_{p+1}+\alpha _{p+1})/L_{p+1},\ldots ,2\pi
(n_{D}+\alpha _{D})/L_{D}),\;n_{p+1},\ldots ,n_{D}=0,\pm 1,\pm 2,\ldots .
\label{kDn}
\end{equation}%
For the components along the uncompactified dimensions one has $-\infty
<k_{l}<\infty $, $l=1,\ldots ,p$.

Substituting the eigenfunctions (\ref{Funcneg}) into formula (\ref{FCsum}),
for the fermionic condensate we find the expression
\begin{equation}
\langle \bar{\psi}\psi \rangle _{p,q}=-\frac{mN}{2^{p+1}\pi ^{p}V_{q}}\int d%
\mathbf{k}_{p}\sum_{\mathbf{n}_{q}\in \mathbf{Z}^{q}}\frac{1}{\omega },
\label{Fermcond}
\end{equation}%
with $\mathbf{n}_{q}=(n_{p+1},\ldots ,n_{D})$ and
\begin{equation}
\omega ^{2}=\mathbf{k}_{p}^{2}+\sum_{l=p+1}^{D}[2\pi (n_{l}+\alpha
_{l})/L_{l}]^{2}+m^{2}.  \label{omega}
\end{equation}%
We implicitly assume the presence of a cutoff function in (\ref{Fermcond})
which makes the inegrosum finite.

For the further evaluation of formula (\ref{Fermcond}) we apply to the sum
over $n_{p+1}$ the Abel-Plana summation formula in the form \cite{Inui03}%
\begin{equation}
\sum_{n_{p+1}=-\infty }^{+\infty }f(|n_{p+1}+\alpha
_{p+1}|)=2\int_{0}^{\infty }dx\,f(x)+i\int_{0}^{\infty }dx\,\sum_{\lambda
=\pm 1}\frac{f(ix)-f(-ix)}{e^{2\pi (x+i\lambda \alpha _{p+1})}-1}.
\label{sumform1}
\end{equation}%
As a result, the fermionic condensate is presented in the decomposed form%
\begin{equation}
\langle \bar{\psi}\psi \rangle _{p,q}=\langle \bar{\psi}\psi \rangle
_{p+1,q-1}+\Delta _{p+1}\langle \bar{\psi}\psi \rangle _{p,q},
\label{FCdecomp}
\end{equation}%
where $\langle \bar{\psi}\psi \rangle _{p+1,q-1}$ corresponds to the first
term on the right-hand side of (\ref{sumform1}) and is the fermionic
condensate for the topology $\mathrm{R}^{p+1}\times (\mathrm{S}^{1})^{q-1}$.
The second term on the right-hand side of formula (\ref{FCdecomp}) is
induced by the compactness of the $z^{p+1}$ direction and is given by the
formula%
\begin{equation}
\Delta _{p+1}\langle \bar{\psi}\psi \rangle _{p,q}=-\frac{2^{-1-p}mNL_{p+1}}{%
\pi ^{(p+1)/2}\Gamma ((p+1)/2)V_{q}}\sum_{\mathbf{n}_{q-1}\in \mathbf{Z}%
^{q-1}}\sum_{\lambda =\pm 1}\int_{\omega _{\mathbf{n}_{q-1}}}^{\infty }du\,%
\frac{(u^{2}-\omega _{\mathbf{n}_{q-1}}^{2})^{(p-1)/2}}{e^{L_{p+1}u+2\pi
i\lambda \alpha _{p+1}}-1},  \label{DeltaFC}
\end{equation}%
where $\mathbf{n}_{q-1}=(n_{p+2},\ldots ,n_{D})$ and%
\begin{equation}
\omega _{\mathbf{n}_{q-1}}^{2}=\sum_{l=p+2}^{D}[2\pi (n_{l}+\alpha
_{l})/L_{l}]^{2}+m^{2}.  \label{omeganq}
\end{equation}%
Note that the expression on the right-hand side of (\ref{DeltaFC}) is finite
and the introduction of the cutoff function is necessary in the first term
on the right of (\ref{FCdecomp}) only.

Expanding the function $1/(e^{y}-1)$ in the integrand of formula (\ref%
{DeltaFC}), we find an alternative form
\begin{equation}
\Delta _{p+1}\langle \bar{\psi}\psi \rangle _{p,q}=-\frac{2NmL_{p+1}}{(2\pi
)^{p/2+1}V_{q}}\sum_{n=1}^{\infty }\cos (2\pi n\alpha _{p+1})\sum_{\mathbf{n}%
_{q-1}\in \mathbf{Z}^{q-1}}\omega _{\mathbf{n}_{q-1}}^{p}f_{p/2}(nL_{p+1}%
\omega _{\mathbf{n}_{q-1}}).  \label{FC}
\end{equation}%
with the notation $f_{\nu }(x)=K_{\nu }(x)/x^{\nu }$. From here it follows
that in the case of periodic boundary condition along the direction $z^{p+1}$
($\alpha _{p+1}=0$) the contribution to the fermionic condensate due to the
compactness of the corresponding direction is always negative independently
of the boundary conditions along the other directions. In the limit when the
length of one of the compactified dimensions, say $z^{l}$, $l\geqslant p+2$,
is large, the main contribution into the sum over $n_{l}$ in (\ref{FC})
comes from large values of $n_{l}$, and in the leading order we can replace
the summation by the integration in accordance with
\begin{equation*}
\frac{1}{L_{l}}\sum_{n_{l}=-\infty }^{+\infty }f(2\pi |n_{l}+\alpha
_{l}|/L_{l})\rightarrow \frac{1}{\pi }\int_{0}^{\infty }dy\,f(y).
\end{equation*}%
The integral over $y$ is evaluated by using the formula
\begin{equation}
\frac{1}{\pi }\int_{0}^{\infty }dy(y^{2}+b^{2})^{p/2}f_{p/2}(c\sqrt{%
y^{2}+b^{2}})=\frac{b^{p+1}}{\sqrt{2\pi }}f_{(p+1)/2}(cb),  \label{IntForm3}
\end{equation}%
and from (\ref{FC}) the corresponding formula is obtained for the topology $%
R^{p+1}\times (S^{1})^{q-1}$. In the limit $L_{l}\ll L_{p+1}$, $l=p+2,\ldots
,D$, the main contribution into the topological part (\ref{FC}) comes from
the term with $\mathbf{n}_{q-1}=0$ and in the leading order we have
\begin{equation}
\Delta _{p+1}\langle \bar{\psi}\psi \rangle _{p,q}\approx -\frac{%
2Nm^{p+1}L_{p+1}}{(2\pi )^{p/2+1}V_{q}}\sum_{n=1}^{\infty }\cos (2\pi
n\alpha _{p+1})f_{p/2}(nL_{p+1}m).  \label{smallLl}
\end{equation}%
As we could expect, for large masses, $mL_{p+1}\gg 1$, the fermionic
condensate given by formula (\ref{FC}) is exponentially suppressed.

After the recurring application of formula (\ref{FC}), the topological part
of the fermionic condensate for spatial topology $\mathrm{R}^{p}\times (%
\mathrm{S}^{1})^{q}$ is presented in the form%
\begin{equation}
\langle \bar{\psi}\psi \rangle _{p,q}=\sum_{j=p}^{D-1}\Delta _{j+1}\langle
\bar{\psi}\psi \rangle _{j,D-j}.  \label{FCTop}
\end{equation}%
For a massless field the fermionic condensate vanishes.

An alternative form for the topological part in the fermionic condensate is
obtained by making use of the zeta function technique \cite{Eliz94,Kirs01}.
We introduce the zeta function density%
\begin{equation}
\zeta (s)=\frac{1}{V_{q}}\int \frac{d\mathbf{k}_{p}}{(2\pi )^{p}}\sum_{%
\mathbf{n}_{q}\in \mathbf{Z}^{q}}\frac{1}{\omega ^{2s}},  \label{zeta}
\end{equation}%
with $\omega $ defined by relation (\ref{omega}). In the case $\alpha _{l}=0$%
, $m=0$, the point $\mathbf{n}_{q}=0$ is to be excluded from the sum. After
the integration over $\mathbf{k}_{p}$, this function is presented in the form%
\begin{equation}
\zeta (s)=\frac{\Gamma (s-p/2)}{(4\pi )^{p/2}\Gamma (s)V_{q}}\sum_{\mathbf{n}%
_{q}\in \mathbf{Z}^{q}}\left\{ \sum_{l=p+1}^{D}[2\pi (n_{l}+\alpha
_{l})/L_{l}]^{2}+m^{2}\right\} ^{p/2-s}.  \label{zeta1}
\end{equation}

An exponentially convergent expression for the analytic continuation of the
function (\ref{zeta1}) is given by the generalized Chowla-Selberg formula
\cite{Eliz98}. The application of this formula to Eq. (\ref{zeta1}) gives%
\begin{equation}
\zeta (s)=\zeta _{\mathrm{M}}(s)+\zeta _{p,q}(s),  \label{zetadec}
\end{equation}%
where%
\begin{equation}
\zeta _{\mathrm{M}}(s)=\int \frac{d\mathbf{k}_{D}}{(2\pi )^{D}}\frac{1}{%
(k_{D}^{2}+m^{2})^{s}}=\frac{m^{D-2s}}{(4\pi )^{D/2}}\frac{\Gamma (s-D/2)}{%
\Gamma (s)},  \label{zetaM}
\end{equation}%
is the corresponding zeta function in the usual Minkowski spacetime and the
part
\begin{equation}
\zeta _{p,q}(s)=\frac{2^{1-s}m^{D-2s}}{(2\pi )^{D/2}\Gamma (s)}%
\sideset{}{'}{\sum}_{\mathbf{m}_{q}\in \mathbf{Z}^{q}}\cos (2\pi \mathbf{m}%
_{q}\cdot \boldsymbol{\alpha }_{q})f_{D/2-s}(mg(\mathbf{L}_{q},\mathbf{m}%
_{q})),  \label{zeta2}
\end{equation}%
with $\mathbf{L}_{q}=(L_{p+1},\ldots ,L_{D})$ and $\boldsymbol{\alpha }%
_{q}=(\alpha _{p+1},\ldots ,\alpha _{D})$, is induced by the nontrivial
topology. The prime on the summation sign in (\ref{zeta2}) means that the
term $\mathbf{m}_{q}=0$ should be excluded from the sum and we have used the
notation%
\begin{equation}
g(\mathbf{L}_{q},\mathbf{m}_{q})=\left(
\sum_{i=p+1}^{D}L_{i}^{2}m_{i}^{2}\right) ^{1/2}.  \label{gLm}
\end{equation}

The topological part in (\ref{zetadec}) is an analytic function at the
physical point $s=1/2$ and for the fermionic condensate one directly finds%
\begin{equation}
\langle \bar{\psi}\psi \rangle _{p,q}=-\frac{mN}{2}\zeta _{p,q}(1/2)=-\frac{%
Nm^{D}}{(2\pi )^{(D+1)/2}}\sideset{}{'}{\sum}_{\mathbf{m}_{q}\in \mathbf{Z}%
^{q}}\cos (2\pi \mathbf{m}_{q}\cdot \boldsymbol{\alpha }_{q})f_{(D-1)/2}(mg(%
\mathbf{L}_{q},\mathbf{m}_{q})).  \label{FCTop2}
\end{equation}%
In the case $p=D-1$, $q=1$ this formula coincides with (\ref{FC}). In
appendix we prove the equivalence of two representations (\ref{FCTop}) and (%
\ref{FCTop2}) for the topological part in the fermionic condensate for
general case. Note that in (\ref{FCTop2}) we can write the function $\cos
(2\pi \mathbf{m}_{q}\cdot \boldsymbol{\alpha }_{q})$ in the form of the
product $\prod_{i=p+1}^{D}\cos (2\pi m_{i}\alpha _{i})$.

\section{Energy-momentum tensor}

\label{sec:EMT}

In order to find the VEV for the operator of the energy-momentum tensor, we
substitute the expansion (\ref{operatorexp}) and the analog expansion for
the operator $\hat{\bar{\psi}}$ into the corresponding expression for spinor
fields,
\begin{equation}
T_{\mu \nu }\{\hat{\bar{\psi}},\hat{\psi}\}=\frac{i}{2}[\hat{\bar{\psi}}%
\gamma _{(\mu }\partial _{\nu )}\hat{\psi}-(\partial _{(\mu }\hat{\bar{\psi}}%
)\gamma _{\nu )}\hat{\psi}]\ .  \label{EMTform}
\end{equation}%
Similar to the case of the fermionic condensate, by making use of the
commutation relations for the annihilation and creation operators, one finds
the following mode-sum formula
\begin{equation}
\langle 0|T_{\mu \nu }|0\rangle =\langle T_{\mu \nu }\rangle
_{p,q}=\sum_{\beta }T_{\mu \nu }\{\bar{\psi}_{\beta }^{(-)}(x),\psi _{\beta
}^{(-)}(x)\}\ .  \label{modesum}
\end{equation}%
Substituting the eigenfunctions (\ref{Funcneg}) into this mode-sum formula,
for the energy density and vacuum stresses one finds (no summation over $%
l=1,\ldots ,D$)%
\begin{eqnarray}
\langle T_{0}^{0}\rangle _{p,q} &=&-\frac{N}{2(2\pi )^{p}V_{q}}\int d\mathbf{%
k}_{p}\sum_{\mathbf{n}_{q}\in \mathbf{Z}^{q}}\omega ,  \label{T00} \\
\langle T_{l}^{l}\rangle _{p,q} &=&\frac{N}{2(2\pi )^{p}V_{q}}\int d\mathbf{k%
}_{p}\sum_{\mathbf{n}_{q}\in \mathbf{Z}^{q}}\frac{k_{l}^{2}}{\omega }.
\label{Tll}
\end{eqnarray}%
As in the case of the fermionic condensate, we will assume that some cutoff
function is present, without writing it explicitly.

After the application of summation formula (\ref{sumform1}) to the series
over $n_{p+1}$, we receive the following recurrence relation%
\begin{equation}
\langle T_{\mu }^{\nu }\rangle _{p,q}=\langle T_{\mu }^{\nu }\rangle
_{p+1,q-1}+\Delta _{p+1}\langle T_{\mu }^{\nu }\rangle _{p,q},
\label{TmunuDecomp}
\end{equation}%
where $\langle T_{\mu }^{\nu }\rangle _{p+1,q-1}$ is the VEV of the
energy-momentum tensor for the topology $R^{p+1}\times (S^{1})^{q-1}$. The
part $\Delta _{p+1}\langle T_{\mu }^{\nu }\rangle _{p,q}$ is induced by the
compactness of the $z^{p+1}$ direction and is given by the expression (no
summation over $l$)%
\begin{equation}
\Delta _{p+1}\langle T_{l}^{l}\rangle _{p,q}=\frac{(4\pi )^{-(p+1)/2}NL_{p+1}%
}{\Gamma ((p+1)/2)V_{q}}\sum_{\mathbf{n}_{q-1}\in \mathbf{Z}%
^{q-1}}\sum_{\lambda =\pm 1}\int_{\omega _{\mathbf{n}_{q-1}}}^{\infty }du\,%
\frac{f^{(l)}(u)(u^{2}-\omega _{\mathbf{n}_{q-1}}^{2})^{(p-1)/2}}{%
e^{L_{p+1}u+2\pi i\lambda \alpha _{p+1}}-1},  \label{Tll2}
\end{equation}%
with the notations%
\begin{eqnarray}
f^{(l)}(u) &=&\frac{4(u^{2}-\omega _{\mathbf{n}_{q-1}}^{2})}{p+1}%
,\;l=0,1,\ldots ,p,  \notag \\
f^{(p+1)}(u) &=&-2u^{2},\;f^{(l)}(u)=k_{l}^{2},\;l=p+2,\ldots D.  \label{flu}
\end{eqnarray}%
Expanding the integrand, this expression can also be presented in the form
(no summation over $l$)%
\begin{equation}
\Delta _{p+1}\langle T_{l}^{l}\rangle _{p,q}=\frac{2NL_{p+1}}{(2\pi
)^{p/2+1}V_{q}}\sum_{\mathbf{n}_{q-1}\in \mathbf{Z}^{q-1}}\sum_{n=1}^{\infty
}\cos (2\pi n\alpha _{p+1})\omega _{\mathbf{n}_{q-1}}^{p+2}F^{(l)}(nL_{p+1}%
\omega _{\mathbf{n}_{q-1}}),  \label{Tll3}
\end{equation}%
with the notations%
\begin{eqnarray}
F^{(0)}(z) &=&F^{(l)}(z)=f_{p/2+1}(z),\;l=1,\ldots ,p,  \notag \\
F^{(p+1)}(z) &=&-f_{p/2}(z)-(p+1)f_{p/2+1}(z),  \label{Flu} \\
F^{(l)}(z) &=&(k_{l}/\omega _{\mathbf{n}_{q-1}})^{2}f_{p/2}(z),\;l=p+2,%
\ldots ,D.  \notag
\end{eqnarray}%
It is easy to check that for a massless field the topological part (\ref%
{Tll3}) is traceless. As we see the vacuum stresses along the uncompactified
dimensions are equal to the energy density. Of course, this property is a
direct consequence of the boost invariance along the corresponding
directions. In particular, from (\ref{Tll3}) it follows that in the case of
periodic boundary conditions along the coordinate $z^{p+1}$ ($\alpha
_{p+1}=0 $), the compactification along this coordinate increases the vacuum
energy density independently of the boundary conditions along the other
directions. The limiting cases of general formulae for the VEV of the
energy-momentum tensor are investigated in a way similar to that described
before for the condensate.

From (\ref{TmunuDecomp}), for the VEV of the energy-momentum tensor in the
topology $R^{p}\times (S^{1})^{q}$ one finds
\begin{equation}
\langle T_{\mu }^{\nu }\rangle _{p,q}=\sum_{j=p}^{D-1}\Delta _{j+1}\langle
T_{\mu }^{\nu }\rangle _{j,D-j}.  \label{Tmunutot}
\end{equation}%
Now, by using the standard relations for the Mac-Donald function, it can be
seen that the vacuum energy density and stresses along the compactified
dimensions are related by the formula (no summation over $l$)%
\begin{equation}
\partial _{L_{l}}(V_{q}\langle T_{0}^{0}\rangle _{p,q})=\frac{V_{q}}{L_{l}}%
\langle T_{l}^{l}\rangle _{p,q},\;l=p+1,\ldots ,D.  \label{EnergyStress}
\end{equation}

\begin{figure}[tbph]
\begin{center}
\begin{tabular}{cc}
\epsfig{figure=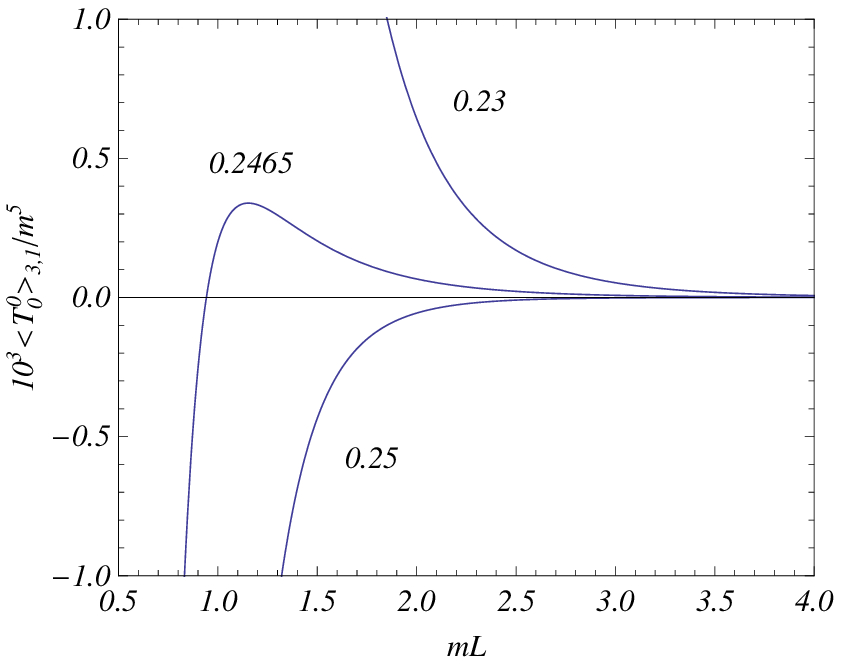,width=7.cm,height=6.cm} & \quad %
\epsfig{figure=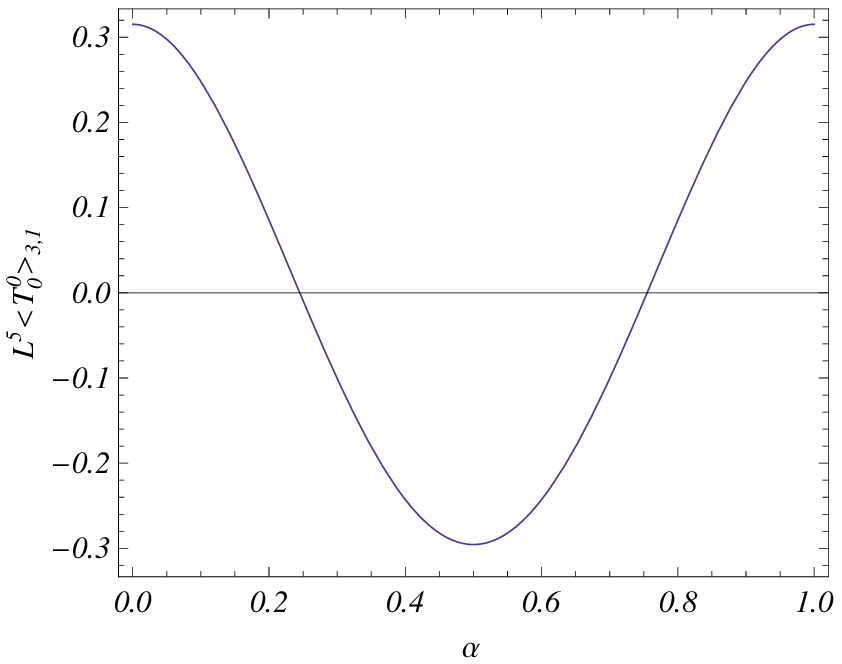,width=7.cm,height=6cm}%
\end{tabular}%
\end{center}
\caption{The Casimir energy density in the Kaluza-Klein-type model with
spatial topology $R^{3}\times S^{1}$ as a function of the parameter $mL$ for
different values of $\protect\alpha $ (left panel). The right panel presents
the corresponding quantity for a massless field as a function of $\protect%
\alpha $.}
\label{fig1}
\end{figure}

For the simplest Kaluza-Klein-type model with spatial topology
$R^{3}\times
S^{1}$, from (\ref{Tll3}) for the energy density one finds ($L_{p+1}=L$, $%
\alpha _{p+1}=\alpha $)
\begin{equation}
\langle T_{0}^{0}\rangle _{3,1}=\frac{1}{\pi ^{2}L^{5}}\sum_{n=1}^{\infty }%
\frac{\cos (2\pi n\alpha )}{n^{5}e^{nmL}}[(nmL)^{2}+3nmL+3].  \label{T00R3S1}
\end{equation}%
This quantity is positive for an untwisted field ($\alpha =0$) and is
negative for a twisted field ($\alpha =1/2$). In the general case, the
Casimir energy density is not a monotonic function of the size of the
compactified dimension. This is seen from the left panel of figure \ref{fig1}
where we have plotted the quantity (\ref{T00R3S1}) as a function of the
parameter $mL$ for different values of the phase $\alpha $ (numbers near the
curves). The values of the phase are chosen in a way to show the transition
from the positive energies to negative ones. In the right panel of figure %
\ref{fig1} we have presented the Casimir energy density (\ref{T00R3S1}) for
a massless field as a function of the parameter $\alpha $.

An alternative expression for the VEV of the energy density is obtained by
using the integral representation of the corresponding zeta function given
by (\ref{zeta2}):%
\begin{equation}
\langle T_{0}^{0}\rangle _{p,q}=-\frac{N}{2}\zeta _{p,q}(-1/2)=\frac{Nm^{D+1}%
}{(2\pi )^{(D+1)/2}}\sideset{}{'}{\sum}_{\mathbf{m}_{q}\in \mathbf{Z}%
^{q}}\cos (2\pi \mathbf{m}_{q}\cdot \boldsymbol{\alpha }_{q})f_{(D+1)/2}(mg(%
\mathbf{L}_{q},\mathbf{m}_{q})).  \label{T00zeta}
\end{equation}%
The equivalence of the representations (\ref{Tmunutot}) and (\ref{T00zeta})
for the energy density is seen in a way similar to that used in appendix for
the fermionic condensate. The corresponding formulae for the vacuum stresses
along compactified dimensions are obtained from relations (\ref{EnergyStress}%
) (no summation over $l$):%
\begin{equation}
\langle T_{l}^{l}\rangle _{p,q}=\langle T_{0}^{0}\rangle _{p,q}-\frac{%
Nm^{D+3}L_{l}^{2}}{(2\pi )^{(D+1)/2}}\sideset{}{'}{\sum}_{\mathbf{m}_{q}\in
\mathbf{Z}^{q}}m_{l}^{2}\cos (2\pi \mathbf{m}_{q}\cdot \boldsymbol{\alpha }%
_{q})f_{(D+3)/2}(mg(\mathbf{L}_{q},\mathbf{m}_{q})),  \label{Tllzeta}
\end{equation}%
with $l=p+1,\ldots ,D$. A number of special cases of formula (\ref{T00zeta})
for the Casimir energy can be found in literature (see \cite{Gonc85},\cite%
{Most97}-\cite{Duff86}, \cite{CasTor}). For a massless fermionic field from (%
\ref{T00zeta}) we find (no summation over $l$)%
\begin{eqnarray}
\langle T_{0}^{0}\rangle _{p,q} &=&N\frac{\Gamma ((D+1)/2)}{2\pi ^{(D+1)/2}}%
\sideset{}{'}{\sum}_{\mathbf{m}_{q}\in \mathbf{Z}^{q}}\frac{\cos (2\pi
\mathbf{m}_{q}\cdot \boldsymbol{\alpha }_{q})}{g^{D+1}(\mathbf{L}_{q},%
\mathbf{m}_{q})},  \label{T00zetam0} \\
\langle T_{l}^{l}\rangle _{p,q} &=&\langle T_{0}^{0}\rangle _{p,q}-N(D+1)%
\frac{\Gamma ((D+1)/2)}{2\pi ^{(D+1)/2}}\sideset{}{'}{\sum}_{\mathbf{m}%
_{q}\in \mathbf{Z}^{q}}L_{l}^{2}m_{l}^{2}\frac{\cos (2\pi \mathbf{m}%
_{q}\cdot \boldsymbol{\alpha }_{q})}{g^{D+3}(\mathbf{L}_{q},\mathbf{m}_{q})},
\label{Tllzeta0}
\end{eqnarray}%
where $l=p+1,\ldots ,D$. Note that for a massless field the representation (%
\ref{Tll3}) has stronger convergence than the one given by (\ref{T00zetam0}%
), (\ref{Tllzeta0}): the summand in (\ref{Tll3}) decays exponentially
instead of power-law decay in (\ref{T00zetam0}), (\ref{Tllzeta0}).

\section{Applications to nanotubes}

\label{sec:Nanotubes}

In this section we specify the general results given above for the electrons
on a carbon sheet rolled into a cylinder or torus making use of the
description of the electronic states in terms of Dirac fermion fields. In
this case $D=2$ and we consider the geometries of cylindrical and toroidal
nanotubes separately. Note that the Dirac-like model for electrons in a
carbon nanotube is valid provided that the cylinder circumference is much
larger than the interatomic spacing. For typical nanotubes the corresponding
ratio can be between 10 and 20 and this approximation is adequate \cite%
{Sait98,Seme84}.

\subsection{Cylindrical nanotubes}

A single wall cylindrical nanotube is a graphene sheet rolled into a
cylindrical shape. For this case we have spatial topology $R^{1}\times S^{1}$
with the compactified dimension of the length $L$. Note that the carbon
nanotube is characterized by its chiral vector $\mathbf{C}_{h}=n_{w}\mathbf{a%
}_{1}+m_{w}\mathbf{a}_{2}$, with $n_{w}$, $m_{w}$ being integers, and $L=|%
\mathbf{C}_{h}|=a\sqrt{n_{w}^{2}+m_{w}^{2}+n_{w}m_{w}}$. In the expression
for the chiral vector, $\mathbf{a}_{1}$ and $\mathbf{a}_{2}$ are the basis
vectors of the hexagonal lattice of graphene and $a=|\mathbf{a}_{1}|=|%
\mathbf{a}_{2}|=2.46\mathring{A}$ is the lattice constant. A zigzag nanotube
corresponds to the special case $\mathbf{C}_{h}=(n_{w},0)$, and a armchair
nanotube corresponds to the case $\mathbf{C}_{h}=(n_{w},n_{w})$. All other
cases correspond to chiral nanotubes. The electron properties of carbon
nanotubes can be either metallic or semiconductor like depending on the
manner the cylinder is obtained from the graphene sheet. In the case $%
n_{w}-m_{w}=3q_{w}$, $q_{w}\in Z$, the nanotube will be metallic
and in the case $n_{w}-m_{w}\neq 3q_{w}$ the nanotube will be
semiconductor with an energy gap inversely proportional to the
diameter. In particular, the armchair nanotube is metallic and the
$(n_{w},0)$ zigzag nanotube is metallic if and only if $n_{w}$ is
an integer multiple of 3.

In order to see the boundary conditions along the compactified dimension, we
note that for the $(n_{w},m_{w})$ nanotube the phase factor in the
wavefunction is in the form $e^{i[m_{1}+(n_{w}-m_{w})/3]\varphi }$, $%
m_{1}\in Z$, where $\varphi $ is the angular variable along the compact
dimension. From here it follows that for metallic nanotubes we have periodic
boundary conditions ($\alpha _{l}=0$) and for semiconductor nanotubes,
depending on the chiral vector, we have two classes of inequivalent boundary
conditions corresponding to $\alpha _{l}=\pi /3$ ($n_{w}-m_{w}=3q_{w}+2$)
and $\alpha _{l}=2\pi /3$ ($n_{w}-m_{w}=3q_{w}+1$). In the expression for
the Casimir densities the phases $\alpha _{l}$ appear in the form $\cos
(2\pi n\alpha _{l})$ and, hence, the Casimir energy density and stresses are
the same for these two cases.

Using the tight-binding approximation it can be seen that the electronic
band structure close to the Dirac points shows a conical dispersion $E(%
\mathbf{k})=v_{F}|\mathbf{k}|$, where $\mathbf{k}$ is the momentum measured
relatively to the Dirac points and $v_{F}$ represents the Fermi velocity
which plays the role of speed of light. The corresponding low-energy
excitations can be described by a pair of two-component Weyl spinors, which
are composed of the Bloch states residing on the two different sublattices
of the honeycomb lattice of the graphene sheet. The corresponding Fermi
velocity is given by $v_{F}=3ta/2$ ($v_{F}\approx 10^{8}\mathrm{cm/s}$ in
graphene), where $t$ is the nearest neighbor hopping energy. Below, in
specifying the formulae from previous section for the case $D=2$, we
consider a massive spinor field to keep the discussion general. The formulae
for a massless case, appropriate for carbon nanotubes, will be given
separately.

In the case $D=2$, the general formula for the fermionic condensate from
section \ref{sec:WF} takes the form ($N=2$, $p=1$,$q=1$,$V_{q}=L$, $%
L_{p+1}\equiv L$, $\alpha _{p+1}\equiv \alpha $)%
\begin{equation}
\langle \bar{\psi}\psi \rangle _{1,1}=-\frac{m}{\pi L}S_{\alpha }(mL),
\label{FCNano}
\end{equation}%
where we have defined%
\begin{eqnarray}
S_{\alpha }(x) &=&\sum_{n=1}^{+\infty }\cos (2\pi n\alpha )\frac{e^{-xn}}{n}
\notag \\
&=&-\frac{1}{2}\ln [1-2e^{-x}\cos (2\pi \alpha )+e^{-2x}].  \label{Salfax}
\end{eqnarray}

\begin{figure}[tbph]
\begin{center}
\epsfig{figure=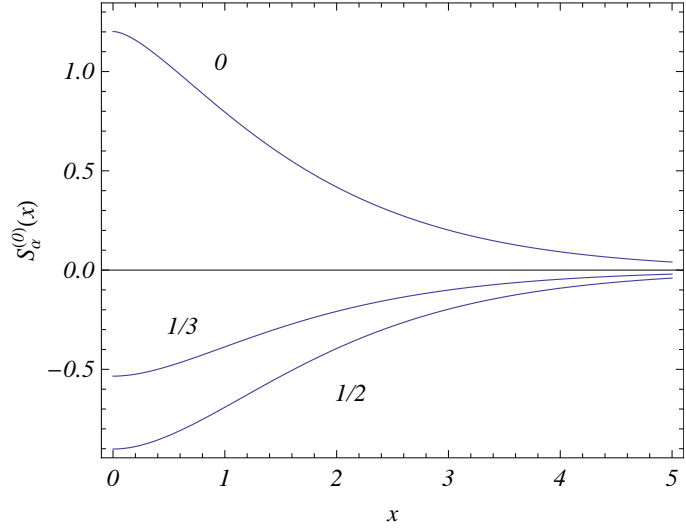,width=9.cm,height=7.cm}
\end{center}
\caption{The function $S_{\protect\alpha }^{(0)}(x)$ from (\protect\ref%
{Salf0}) for different values of the parameter $\protect\alpha $ (numbers
near the curves).}
\label{fig2}
\end{figure}

In a similar way, for the VEV of the energy-momentum tensor from (\ref{Tll3}%
) we find (no summation over $l$)
\begin{equation}
\langle T_{l}^{l}\rangle _{1,1}=\frac{1}{\pi L^{3}}\sum_{n=1}^{\infty }\cos
(2\pi n\alpha )G^{(l)}(nmL)\frac{e^{-nmL}}{n^{3}},  \label{EMTNano}
\end{equation}%
with the notations%
\begin{equation}
G^{(0)}(z)=G^{(1)}(z)=1+z,\;G^{(2)}(z)=-(2+2z+z^{2}).  \label{Gl}
\end{equation}%
In particular, for the energy density we have%
\begin{equation}
\langle T_{0}^{0}\rangle _{1,1}=\frac{1}{\pi L^{3}}S_{\alpha }^{(0)}(mL),
\label{T00tube}
\end{equation}%
where the notation%
\begin{equation}
S_{\alpha }^{(0)}(x)=\sum_{n=1}^{\infty }\cos (2\pi n\alpha )e^{-nx}\frac{%
1+nx}{n^{3}},  \label{Salf0}
\end{equation}%
is introduced. In figure \ref{fig2} we have plotted the function $S_{\alpha
}^{(0)}(x)$ for different values of $\alpha $ (numbers near the curves). In
particular, the Casimir energy density is positive for armchair nanotubes
(periodic boundary conditions).

In the case $m=0$ we have%
\begin{equation}
\langle T_{0}^{0}\rangle _{1,1}=\langle T_{1}^{1}\rangle _{1,1}=-\frac{1}{2}%
\langle T_{2}^{2}\rangle _{1,1}=\frac{S_{\alpha }^{(0)}(0)}{\pi L^{3}},
\label{T00m0D2}
\end{equation}%
In particular, $S_{0}^{(0)}(0)=1.202$, $S_{1/2}^{(0)}(0)=-0.902$, and $%
S_{1/3}^{(0)}(0)=-0.534$. Note that the corresponding fermionic condensate
vanishes. In carbon nanotubes we have two sublattices and each of them gives
the contribution to the Casimir densities given by (\ref{T00m0D2}). So, for
the Casimir energy density on a carbon nanotube with radius $L$ one has%
\begin{equation}
\langle T_{0}^{0}\rangle _{1,1}^{\mathrm{(cn)}}=\frac{2\hbar v_{F}}{\pi L^{3}%
}S_{\alpha }^{(0)}(0),  \label{T00cn}
\end{equation}%
where the standard units are restored. Hence, we see that the topological
Casimir energy is positive for metallic nanotubes and is negative for
semiconducting ones.

\subsection{Toroidal nanotubes}

For the geometry of a toroidal nanotube we have the spatial topology $%
(S^{1})^{2}$ with $p=0$ and $q=2$. In this case from the general formulae
for the fermionic condensate we find%
\begin{equation}
\langle \bar{\psi}\psi \rangle _{0,2}=-\frac{m}{\pi }\sum_{j=1,2}\frac{%
S_{\alpha _{j}}(mL_{j})}{L_{j}}-\frac{2m}{\pi }\sum_{m_{1}=1}^{\infty
}\sum_{m_{2}=1}^{\infty }\cos (2\pi m_{1}\alpha _{1})\cos (2\pi m_{2}\alpha
_{2})\frac{e^{-m\sqrt{m_{1}^{2}L_{1}^{2}+m_{2}^{2}L_{2}^{2}}}}{\sqrt{%
m_{1}^{2}L_{1}^{2}+m_{2}^{2}L_{2}^{2}}},  \label{FCD2}
\end{equation}%
where the function $S_{\alpha }(x)$ is defined by (\ref{Salfax}).

For the energy density ant the vacuum stresses the corresponding formulae
have the form (no summation over $l$)%
\begin{eqnarray}
\langle T_{0}^{0}\rangle _{0,2} &=&\sum_{j=1,2}\frac{S_{\alpha
_{j}}^{(0)}(mL_{j})}{\pi L_{j}^{3}}+\frac{2}{\pi }\sum_{m_{1}=1}^{\infty
}\sum_{m_{2}=1}^{+\infty }\frac{\cos (2\pi m_{1}\alpha _{1})\cos (2\pi
m_{2}\alpha _{2})}{\exp (mg(\mathbf{L}_{2},\mathbf{m}_{2}))}\frac{1+mg(%
\mathbf{L}_{2},\mathbf{m}_{2})}{g^{3}(\mathbf{L}_{2},\mathbf{m}_{2})},
\label{T00D2tor} \\
\langle T_{l}^{l}\rangle _{0,2} &=&\langle T_{0}^{0}\rangle _{0,2}-\frac{%
m^{5}}{\pi }\sum_{j=1,2}\sum_{m_{j}=1}^{+\infty }\cos (2\pi m_{j}\alpha
_{j})L_{l}^{2}m_{l}^{2}\frac{3+3x+x^{2}}{x^{5}e^{x}}|_{x=mL_{j}m_{j}}  \notag
\\
&&-\frac{2m^{5}}{\pi }\sum_{m_{1}=1}^{+\infty }\sum_{m_{2}=1}^{+\infty }\cos
(2\pi m_{1}\alpha _{1})\cos (2\pi m_{2}\alpha _{2})L_{l}^{2}m_{l}^{2}\frac{%
3+3x+x^{2}}{x^{5}e^{x}}|_{x=mg(\mathbf{L}_{2},\mathbf{m}_{2})},
\label{TllD2tor}
\end{eqnarray}%
with $l=1,2$ and $g(\mathbf{L}_{2},\mathbf{m}_{2})=\sqrt{%
m_{1}^{2}L_{1}^{2}+m_{2}^{2}L_{2}^{2}}$. Alternative expressions for the
topological parts are obtained from formulae (\ref{Tll3}) and (\ref{Tmunutot}%
). For a massless field we find
\begin{eqnarray}
\langle T_{0}^{0}\rangle _{0,2} &=&\sum_{j=1,2}\frac{S_{\alpha _{j}}^{(0)}(0)%
}{\pi L_{j}^{3}}+\frac{2}{\pi }\sum_{m_{1}=1}^{\infty
}\sum_{m_{2}=1}^{+\infty }\frac{\cos (2\pi m_{1}\alpha _{1})\cos (2\pi
m_{2}\alpha _{2})}{(m_{1}^{2}L_{1}^{2}+m_{2}^{2}L_{2}^{2})^{3/2}},
\label{T00D2m0tor} \\
\langle T_{l}^{l}\rangle _{0,2} &=&\langle T_{0}^{0}\rangle _{0,2}-\frac{3}{%
\pi }\sum_{j=1,2}\sum_{m_{j}=1}^{+\infty }\cos (2\pi m_{j}\alpha _{j})\frac{%
L_{l}^{2}m_{l}^{2}}{L_{j}^{5}m_{j}^{5}}  \notag \\
&&-\frac{6}{\pi }\sum_{m_{1}=1}^{+\infty }\sum_{m_{2}=1}^{+\infty
}L_{l}^{2}m_{l}^{2}\frac{\cos (2\pi m_{1}\alpha _{1})\cos (2\pi m_{2}\alpha
_{2})}{(m_{1}^{2}L_{1}^{2}+m_{2}^{2}L_{2}^{2})^{5/2}}.  \label{TllD2m0tor}
\end{eqnarray}

\begin{figure}[tbph]
\begin{center}
\begin{tabular}{cc}
\epsfig{figure=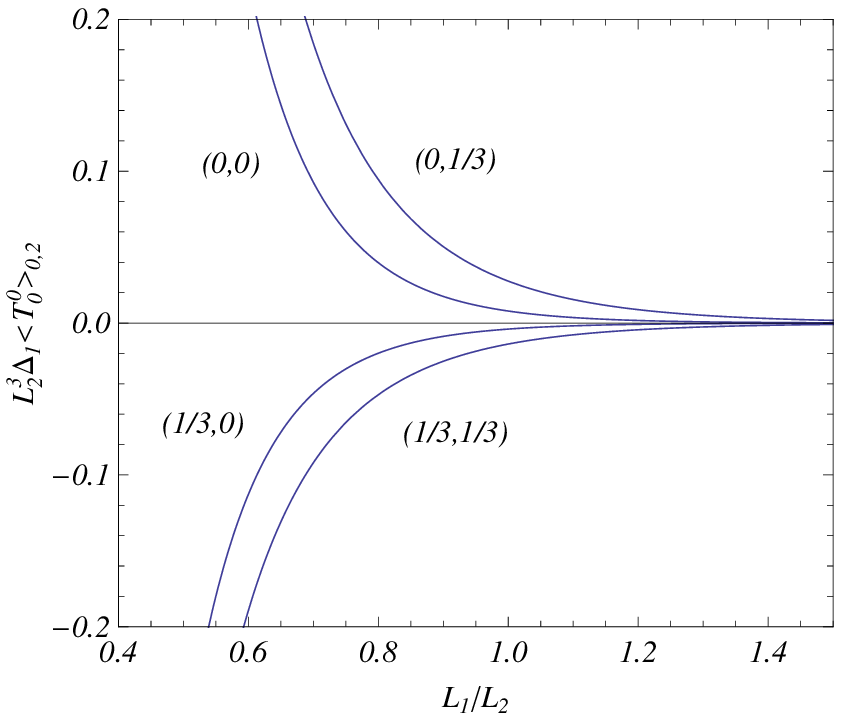,width=7.cm,height=6.cm} & \quad %
\epsfig{figure=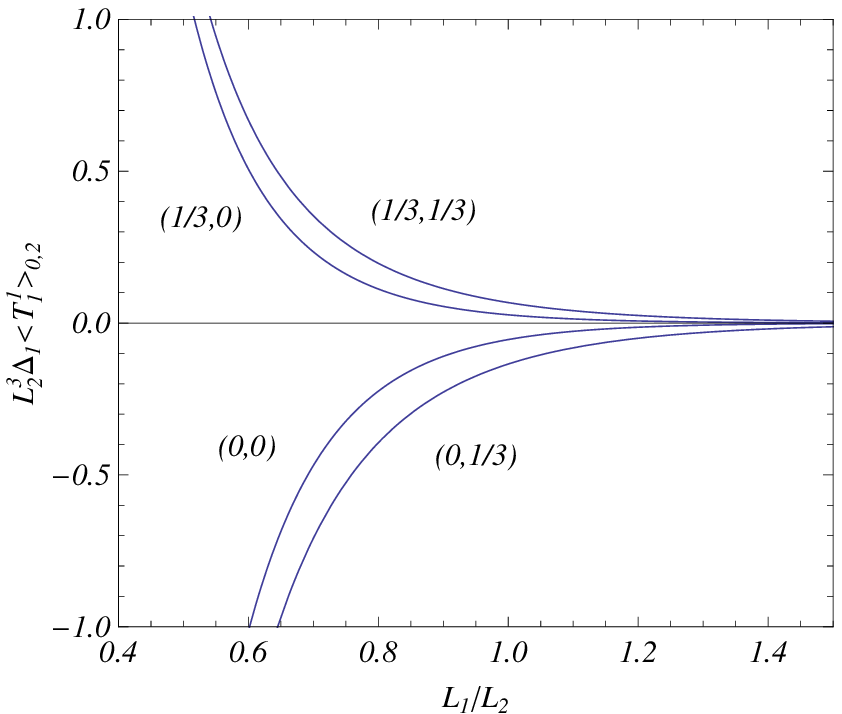,width=7.cm,height=6cm}%
\end{tabular}%
\end{center}
\caption{The difference between the vacuum energy densities (left panel) and
stresses (right panel) between the cylindrical (with radius $L_{2}$) and
toroidal (with radii $L_{1}$ and $L_{2}$) geometries for a massless
fermionic field. The numbers near the curves are the corresponding values
for $(\protect\alpha _{1},\protect\alpha _{2})$.}
\label{fig3}
\end{figure}

In particular, it is of interest to see the difference of the Casimir
densities between the toroidal (with radii $L_{1}$ and $L_{2}$) and
cylindrical (with radius $L_{2}$) geometries of the carbon nanotube. For the
condensate this difference is directly given by formula (\ref{FC}) and one
has
\begin{equation}
\langle \bar{\psi}\psi \rangle _{0,2}=\langle \bar{\psi}\psi \rangle _{1,1}-%
\frac{2m}{\pi L_{2}}\sum_{n=1}^{\infty }\cos (2\pi n\alpha
_{1})\sum_{n_{2}=-\infty }^{+\infty }K_{0}(n(L_{1}/L_{2})\sqrt{4\pi
^{2}(n_{2}+\alpha _{2})^{2}+m^{2}L_{2}^{2}}).  \label{FCD2b}
\end{equation}%
The first term on the right of this formula is the condensate for the
topology $R^{1}\times S^{1}$ with the length of the compactified dimension $%
L_{2}$. Similar formula for the VEV of the energy-momentum tensor follows
from (\ref{Tll3}) (no summation over $l$):
\begin{eqnarray}
\langle T_{l}^{l}\rangle _{0,2} &=&\langle T_{l}^{l}\rangle _{1,1}+\frac{2}{%
\pi L_{2}^{3}}\sum_{n=1}^{\infty }\cos (2\pi n\alpha
_{1})\sum_{n_{2}=-\infty }^{+\infty }\left[ 4\pi ^{2}(n_{2}+\alpha
_{2})^{2}+m^{2}L_{2}^{2}\right]  \notag \\
&&\times F^{(l)}(n(L_{1}/L_{2})\sqrt{4\pi ^{2}(n_{2}+\alpha
_{2})^{2}+m^{2}L_{2}^{2}}),  \label{DeltaT00D2}
\end{eqnarray}%
where the functions $F^{(l)}(z)$ are given by expressions (\ref{Flu}) with $%
p=0$. The second terms on the right-hand sides of formulae (\ref{FCD2b}) and
(\ref{DeltaT00D2}) are induced by the compactification of the cylinder (with
radius $L_{2}$) along its axis. In figure \ref{fig3} we have plotted these
terms for the energy density, $\Delta _{1}\langle T_{0}^{0}\rangle _{0,2}$
(left panel), and for the stress along the axis of the cylinder, $\Delta
_{1}\langle T_{1}^{1}\rangle _{0,2}$ (right panel), for a massless fermionic
field as functions of the ratio $L_{1}/L_{2}$. The numbers near the curves
correspond to the values of $(\alpha _{1},\alpha _{2})$. As we have
mentioned before the values of the phase $\alpha _{l}=0,1/3$ are realized in
carbon nanotubes. The vacuum stress $\Delta _{1}\langle T_{2}^{2}\rangle
_{0,2}$ is related to the quantities plotted in figure \ref{fig3} by the
zero trace condition for the energy-momentum tensor of a massless field.

The corresponding formulae for the Casimir densities in toroidal nanotubes,
which we denote by $\langle T_{l}^{l}\rangle _{0,2}^{\mathrm{(tn)}}$, are
obtained from (\ref{T00D2m0tor}), (\ref{TllD2m0tor}) and (\ref{DeltaT00D2})
in the massless limit with additional factor 2 which takes into account the
presence of two sublattices: $\langle T_{l}^{l}\rangle _{0,2}^{\mathrm{(tn)}%
}=2\langle T_{l}^{l}\rangle _{0,2}|_{m=0}$. In standard units the factor $%
\hbar v_{F}$ appears as well. Note that if the chiral vector $\mathbf{C}_{h}$
is directed along the axis $z^{2}$ then one has $L_{2}=|\mathbf{C}_{h}|$.
The translational vector defining the unit cell, $\mathbf{T}$, is
perpendicular to $\mathbf{C}_{h}$ and its components are related to the
components of the chiral vector by the formula%
\begin{equation}
\mathbf{T=}\frac{n_{w}+2m_{w}}{d_{r}}\mathbf{a}_{1}-\frac{2n_{w}+m_{w}}{d_{r}%
}\mathbf{a}_{2},  \label{Tvec}
\end{equation}%
where $d_{r}=\gcd (n_{w},m_{w})$ if $(m_{w}-n_{w})$ is not a multiple of $%
3\times \gcd (n_{w},m_{w})$ and $d_{r}=3\times \gcd (n_{w},m_{w})$ if $%
(m_{w}-n_{w})$ is a multiple of $3\times \gcd (n_{w},m_{w})$. Here gcd means
the greatest common divisor. Now for the length of the second toroidal
dimension we have $L_{1}=N_{w}|\mathbf{T}|$, where $N_{w}$ is the number of
unit cells along the corresponding direction. By taking into account that $|%
\mathbf{T}|=\sqrt{3}L_{1}/d_{r}$, for the ratio of the lengths of the torus
in (\ref{DeltaT00D2}) one finds $L_{1}/L_{2}=$ $\sqrt{3}N_{w}/d_{r}$. From
the graphs in figure \ref{fig3} it follows that the toroidal
compactification of a cylindrical nanotube along its axis increases the
Casimir energy for periodic boundary conditions ($\alpha _{1}=0$) and
decreases the Casimir energy for the semiconducting-type compactifications.
In particular, the Casimir energy of the armchair cylindrical nanotube
increases by the compactification if $N_{w}$ is an integer multiple of 3 and
decreases otherwise.

\section{Conclusion}

\label{sec:Conclusion}

In the present paper we have investigated the topological Casimir effect for
a massive spinor field on background of spacetime with an arbitrary number
toroidally compactified spatial dimensions. The boundary conditions along
compactified dimensions are taken in general form with arbitrary phases. For
the evaluation of the Casimir densities we have used the direct
mode-summation method. By applying to the corresponding mode-sums the
Abel-Plana formula, we have derived recurrence formulae which relate the
VEVs for the topologies $R^{p}\times (S^{1})^{q}$ and $R^{p+1}\times
(S^{1})^{q-1}$. The part induced by the compactness of the $(p+1)$-th
direction is given by expression (\ref{FC}) for the fermionic condensate and
by expression (\ref{Tll3}) for the VEV\ of the energy-momentum tensor. The
total topological VEVs are obtained after the summation over all
compactified dimensions, formulae (\ref{FCTop}) and (\ref{Tmunutot}).
Alternative expressions are obtained by using the generalized Chowla-Selberg
formula for the analytic continuation of the corresponding zeta function.
These expressions are given by formula (\ref{FCTop2}) for the condensate and
by formulae (\ref{T00zeta}) and (\ref{Tllzeta}) for the energy density and
vacuum stresses along compactified dimensions. Note that the stresses along
the uncompactified dimensions coincide with the energy density. This
property is a direct consequence of the boost invariance along the
corresponding directions. For a massless fermionic field the condensate
vanishes and the expressions for the VEVs of the energy density and vacuum
stresses take the form (\ref{T00zetam0}) and (\ref{Tllzeta0}). Note that,
unlike to the case of a massive field, the convergence of the multiseries in
the latter case is power-law. In the representation based on the application
of the Abel-Plana summation formula we have exponentially convergent
multiseries in both cases of massive and massless fields. On the example of
the simplest Kaluza-Klein-type model with spatial topology $R^{3}\times
S^{1} $ we have demonstrated that, unlike to the special cases of twisted
and untwisted fields, in general, the Casimir energy density is not a
monotonic function of the size of the internal space.

In section \ref{sec:Nanotubes} we specify the general formulae for the model
with $D=2$. This model may be used for the evaluation of the Casimir
densities within the framework of the Dirac-like theory for the description
of the electronic states in carbon nanotubes where the role of speed of
light is played by the Fermi velocity. Though the corresponding spinor field
is massless, to keep the discussion general, we present the formulae for the
cylindrical and toroidal geometries in the massive case and specify the
results for the nanotubes separately. For carbon nanotubes the fermionic
condensate vanishes and the VEV of the energy-momentum tensor is given by
formulae (\ref{T00cn}) for cylindrical nanotubes and by (\ref{T00D2m0tor})
and (\ref{TllD2m0tor}) (with an additional factor 2 which takes into account
the presence of two sublattices) for toroidal nanotubes. In the case of
toroidal nanotubes an alternative representation with the stronger
convergence of the series is given by formula (\ref{DeltaT00D2}) with $m=0$.
The topological Casimir energy is positive for metallic cylindrical
nanotubes and is negative for semiconducting ones. We have shown that the
toroidal compactification of a cylindrical nanotube along its axis increases
the Casimir energy for periodic boundary conditions and decreases the
Casimir energy for the semiconducting-type compactifications. In particular,
the Casimir energy of the armchair cylindrical nanotube increases by the
compactification if the number of unit cells along the axis of cylinder is
an integer multiple of 3 and decreases otherwise.

\section*{Acknowledgments}

A.A.S. was supported by the Armenian Ministry of Education and Science Grant
No. 119. A.A.S. gratefully acknowledges the hospitality of the Abdus Salam
International Centre for Theoretical Physics (Trieste, Italy) where part of
this work was done.

\appendix

\section{Equivalence of two approaches}

\label{sec:Appendix}

In this section we show that the formulae (\ref{FCTop}) and (\ref{FCTop2})
for the topological part in the fermionic condensate are equivalent. First
of all we note that from formula (\ref{FCTop2}) one has
\begin{eqnarray}
\langle \bar{\psi}\psi \rangle _{p,q}&=&\langle \bar{\psi}\psi \rangle
_{p+1,q-1}-\frac{2Nm^{D}}{(2\pi )^{(D+1)/2}}\sum_{m_{p+1}=1}^{\infty }\cos
(2\pi m_{p+1}\alpha _{p+1})  \notag \\
&& \times \sum_{\mathbf{m}_{q-1}\in \mathbf{Z}^{q-1}}\cos (2\pi \mathbf{m}%
_{q-1}\cdot \boldsymbol{\alpha }_{q-1})f_{(D-1)/2}(mg(\mathbf{L}_{q},\mathbf{%
m}_{q})).  \label{RecRelZeta}
\end{eqnarray}%
Hence, we should prove the relation%
\begin{eqnarray}
\sum_{\mathbf{m}_{q-1}\in \mathbf{Z}^{q-1}}\cos (2\pi \mathbf{m}_{q-1}\cdot %
\boldsymbol{\alpha }_{q-1})f_{(D-1)/2}(mg(\mathbf{L}_{q},\mathbf{m}_{q}))&=&
\frac{(2\pi )^{(q-1)/2}L_{p+1}}{V_{q}m^{D-1}}\sum_{\mathbf{n}_{q-1}\in
\mathbf{Z}^{q-1}}\omega _{\mathbf{n}_{q-1}}^{p}  \notag \\
&& \times f_{p/2}(nL_{p+1}\omega _{\mathbf{n}_{q-1}}).  \label{Rel1}
\end{eqnarray}%
For this we will use the Poisson's resummation formula
\begin{equation}
\sum_{\mathbf{m}_{q-1}\in \mathbf{Z}^{q-1}}F(\mathbf{x})\delta (\mathbf{x}-%
\mathbf{m}_{q-1})=\sum_{\mathbf{n}_{q-1}\in \mathbf{Z}^{q-1}}F(\mathbf{x}%
)e^{2i\pi \mathbf{n}_{q-1}\cdot \mathbf{x}},  \label{PoisResum}
\end{equation}%
for the function%
\begin{equation}
F(\mathbf{x})=\cos (2\pi \mathbf{x}\cdot \boldsymbol{\alpha }%
_{q-1})f_{(D-1)/2}(m\sqrt{g^{2}(\mathbf{L}_{q-1},\mathbf{x}%
)+L_{p+1}^{2}m_{p+1}^{2}}).  \label{toPois}
\end{equation}

After the integration over $\mathbf{x}$ we find%
\begin{eqnarray}
&&\sum_{\mathbf{m}_{q-1}\in \mathbf{Z}^{q-1}}\cos (2\pi \mathbf{m}%
_{q-1}\cdot \boldsymbol{\alpha }_{q-1})f_{(D-1)/2}(mg(\mathbf{L}_{q},\mathbf{%
m}_{q}))  \notag \\
&&\quad =\sum_{\mathbf{n}_{q-1}\in \mathbf{Z}^{q-1}}\int d\mathbf{x}\,\cos
[2\pi \mathbf{x}\cdot (\boldsymbol{\alpha }_{q-1}+\mathbf{n}%
_{q-1})]f_{(D-1)/2}(m\sqrt{g^{2}(\mathbf{L}_{q-1},\mathbf{x}%
)+L_{p+1}^{2}m_{p+1}^{2}}).  \label{Rel2}
\end{eqnarray}%
For the evaluation of the integral on the right hand side we first introduce
a new integration variables in accordance with $y_{i}=x_{i}L_{i}$ and then
introduce spherical coordinates. The integration over the angular
coordinates is expressed in terms of the Bessel function. At the final step
the integral is evaluated by using the formula \cite{Gradshtein}%
\begin{equation}
\int_{0}^{\infty }dyy^{\mu +1}J_{\mu }(by)\,f_{\nu }(c\sqrt{y^{2}+a^{2}})=%
\frac{b^{\mu }}{c^{2\nu }}(b^{2}+c^{2})^{\nu -\mu -1}f_{\nu -\mu -1}(a\sqrt{%
b^{2}+c^{2}}).  \label{IntForm}
\end{equation}%
This leads to the following result%
\begin{eqnarray}
&&\int d\mathbf{x}\,\cos (2\pi \mathbf{x}\cdot (\boldsymbol{\alpha }_{q-1}+%
\mathbf{n}_{q-1}))f_{(D-1)/2}(m\sqrt{g^{2}(\mathbf{L}_{q-1},\mathbf{x}%
)+L_{p+1}^{2}m_{p+1}^{2}})  \notag \\
&&\quad =\frac{(2\pi )^{(q-1)/2}L_{p+1}}{m^{D-1}V_{q}}\omega _{\mathbf{n}%
_{q-1}}^{p}f_{p/2}(m_{p+1}L_{p+1}\omega _{\mathbf{n}_{q-1}}),
\label{IntForm2}
\end{eqnarray}%
where $\omega _{\mathbf{n}_{q-1}}$ is defined by relation (\ref{omega}).
Substituting this relation into (\ref{Rel2}) leads to the result (\ref{Rel1}%
) which proves the equivalence of two expressions for the topological part.

\end{document}